\documentstyle[preprint,floats,aps,epsf,epsfig,prb,youngtab]{revtex}

\newlength{\bxwidth}\bxwidth=2.5 truein

\newlength{\fight}
\newcommand\ltdash{\raise-1.8pt\hbox{$\scriptscriptstyle |$}}
\newcommand \beq  {\begin{equation}}
\newcommand \eeq  {\end{equation}}
\newcommand \bea {\begin{eqnarray} }
\newcommand \eea {\end{eqnarray}}

\newcommand \rarrow{\rightarrow}
\newcommand \dg{^{\dagger}}
\newcommand \si { \sigma}
\newcommand \ra { \rangle}
\newcommand\la{\langle}

\newcommand\largeh{\hbox{\ooalign {\raise2pt\hbox{$\relbar \joinrel \relbar \joinrel \rightharpoonup$}\crcr
{\hbox{$\leftharpoondown\joinrel\relbar \joinrel\relbar$}}
}}}
\begin{document}
\draft
\title{
Supersymmetric Hubbard operators }
\author{  P. Coleman
$^{1}$ , C. P{\'e}pin $^{2}$
and 
J. Hopkinson$^1$ 
}
\address{$^1$ Center for Materials Theory,
Department of Physics and Astronomy, 
Rutgers University, Piscataway, NJ 08854, USA.
}
\address{$^2$ SPhT, L'Orme des Merisiers, CEA-Saclay, 91191 Gif-sur-Yvette, France.
}
\maketitle
\date{\today}
\maketitle
\begin{abstract}
We develop a supersymmetric representation of the Hubbard operator
which unifies the slave boson and slave fermion representations into
a single $U (1)\times SU (1\vert 1)$ gauge theory, a group with
larger symmetry than the product of two $U (1)$ gauge groups. 
These representations of the Hubbard operator can be 
used to incorporate strong 
Hund's interactions in multi-electron atoms as a constraint.
We show how this method
can be combined with the  $SP (N)$ group to yield a locally
supersymmetric large-N formulation of the  $t-J$ model.
\end{abstract}
\vskip 0.1 truein
\newpage


One of the fascinating aspects of strongly correlated materials is
their propensity to develop novel metallic phases in 
situations where  local moments interact
strongly with mobile electrons.  Examples of such situations include
metals near a metal insulator transition, \cite{hitc}
metals at an anti-ferromagnetic 
quantum critical point\cite{mathur} and anti-ferromagnetic heavy fermion
superconductors.\cite{upd2al3}
These discoveries challenge our
understanding of how spin and charge interact at the brink of
magnetism.

Theoretical approaches to these problems are hindered by the
difficulty of capturing the profound transformation in  spin
correlations that develops at the boundary between antiferromagnetism 
and  paramagnetism.
Usually we 
model these features by representing the spin
as a boson in a magnetic phase,\cite{arovas} or as a fermion in a paramagnetic
phase,\cite{coleman84} but by making this choice, the character of 
spin and charge excitations which appear in an approximate field theory  is
restricted  and lacks the flexibility to 
describe the co-existence of 
strong magnetic correlations within a paramagnetic phase.

These considerations have 
motivated the development of new methods to describe the spin and
charge excitations of a strongly correlated material which avoid 
making the choice between a bosonic or fermionic spin.\cite{gan,pepin,georges,ngai}  This
paper attempts to stimulate further progress in this direction by introducing 
a supersymmetric representation of Hubbard operators.\cite{hubbard}  The method used
here is an  extension of the supersymmetric spin representation
introduced by Coleman, P{\'e}pin and Tsvelik\cite{cpt1,cpt2} (CPT).
Remarkably, the supersymmetry in the  CPT spin
representation survives  the introduction of charge degrees of freedom, opening
the method to a wider range of models. 


Hubbard operators\cite{hubbard} provide a way to describe 
atoms in which Coulomb repulsion prevents double-occupancy
of a given orbital. Suppose  $\vert a \rangle \in 
\{ \vert 0\rangle, \vert \sigma\rangle\}$ describes 
a set of atomic  states involving a charged ``hole'' $\vert 0\rangle $  or a 
neutral spin state  $\vert \sigma \ra$ with  spin component
$\sigma \in \{1 \dots  N \}$ which for generality can have one of 
$N$ possible values. 
The Hubbard operators are written
\bea
X_{{a} {b} } = \vert {a} \ra \langle {b} \vert
\eea
where ${a} ,{b}  \in \{0,N\}$, represent an atomic
state with $N$ possible spin configurations.   The operators
$X_{\sigma  \sigma '}$ are bosonic spin operators 
whereas the 
$X_{\sigma 0}$ and $X_{0\sigma }$ are fermionic operators that respectively
create and annihilate a single electron. The spin operators ${X
}_{\sigma \sigma '}$ are the generators of the group $SU (N)$. The additional
operators  expand the group to a supergroup 
$SU(N\vert 1)$\cite{supergroups} that describes the physical spin and charge degrees of freedom 
of the atom. 
These operators satisfy 
a graded Lie algebra 
\bea
[X_{{a} {b}} , X_{{c} {d}}]_{\pm} = \delta_{{b} {c}}
X_{{a} {d}} \pm \delta_{{a} {d}} X_{{c} {b}}.
\eea
where the plus sign is only used for fermionic operators.
The absence of a  Wick's theorem for 
these operators is normally overcome
by factorizing
the fermionic Hubbard operators 
as a product of canonical creation and annihilation operators.
This can be
done by representing the empty state by a ``slave
boson'' and the spin by a fermion \cite{coleman84} or alternatively,
by representing the empty state as a 
``slave fermion'' and the spin by a Schwinger boson.\cite{slavefermion}

We now generalize this approach, introducing 
\begin{eqnarray}\label{spinors}
F_{a }&=&(f_{1},\dots f_{N},\phi )\cr
B_{a }&=&(b_{1},\dots b_{N},\chi  )
\end{eqnarray}
where $b_{{\sigma } }$ and $f_{{\sigma } }$ denote 
a Schwinger boson\cite{arovas}  and Abrikosov
pseudo-fermion\cite{abrikosov} respectively, while
$\phi $ and $\chi $ are slave bosons\cite{coleman84} and fermions\cite{slavefermion} respectively. 
In terms of these operators, the 
supersymmetric representation of the Hubbard operators
is written 
\begin{eqnarray}\label{xdef}
{X}_{{a} {b}} = B_{{a}}^{\dagger}B_{b} + F_{{a}}^{\dagger}F_{b},
\end{eqnarray}
Written out explicitly, this is 
\begin{eqnarray}\label{therep}
X_{\sigma \sigma '} &=& b\dg _{\sigma }b_{\sigma '}+ f\dg _{\sigma
}f_{\sigma '}\cr
X_{\sigma 0}&=& b\dg _{\sigma }\chi + f\dg _{\sigma }\phi , \qquad 
X_{0\sigma }= \chi \dg b _{\sigma } + \phi\dg f _{\sigma }\cr
X_{00}&=& \chi\dg \chi +\phi \dg \phi 
\end{eqnarray}
By summing the slave fermion and slave boson representations 
we are guaranteed that the representation 
satisfies the correct commutation algebra. 
The novelty of our approach lies 
in the two unique constraints which make the representation 
irreducible, which we show to be
\begin{equation}\label{constraint1}
Q = n_{{ b}}
+ n_{{ \phi}} + n_{{f}} + n_{{ \chi}},
\end{equation}
the total number of particles and 
\begin{equation}\label{constraint2}
Y = n_{{\phi}} + n_{{f}} - (n_{{ b}} + n_{{\chi}}) + \frac{1}{Q}[\theta, \theta^{\dagger}],
\end{equation}
the ``asymmetry'' of the representation, where
$\theta = \sum_{\sigma}{b}_{\sigma }^{\dagger}{f}_{\sigma } - {\chi}^{\dagger}{\phi}$  and its conjugate
$\theta\dg$ 
are fermionic operators which satisfy the algebra
$\{\theta,\theta^{\dagger}\} = Q$. 
The $\theta $ operators interconvert bosons and fermions. 
\[
b_{\sigma }{\mathop {\largeh}^{\displaystyle\theta \dg }_{\displaystyle
\theta } }f_{\sigma }, \qquad 
-\chi {\mathop {\largeh }^{\displaystyle \theta \dg }_{\displaystyle
\theta } }\phi  \qquad \nonumber
\]

The special feature of this representation
is that $\theta $ and $\theta \dg $ commute with the constraints
$[\theta^{(\dagger )} ,Q]=[\theta^{(\dagger )}
,{Y}]=0$, 
the bosonic 
Hubbard operators
\[
[\theta^{(\dagger )} , X_{\sigma \sigma'}]= [\theta^{(\dagger )}
,X_{00}] = 
0,
\]
and they also anti-commute with the fermionic Hubbard operators
\[
\{ \theta ^{(\dagger )},X_{\sigma 0}\} = \{\theta^{(\dagger )} ,X_{{0\sigma }} \}=0,
\]
so that there is a {\sl local}
supersymmetry which underlies the constraint. 
The 
operators $Q$, 
$\theta $ and $\theta \dg $ are the
generators of the simplest supergroup  $SU (1\vert 1)$\cite{supergroups}; the 
operator $Y$ generates an additional $U (1)$ symmetry. 
Remarkably, by combining the slave boson and slave fermion
representations, the abelian gauge groups of the starting representation
``fuse'' into a supergroup with 
greater symmetry $U_{SB} (1)\times U_{SF} (1)\rarrow  U (1)\times SU (1\vert 1).$
If we introduce the operator $\hat A = [\bar \eta \theta - \theta
\dg  \eta ]$, where $\eta $ and $\bar \eta $ are Grassman numbers, 
then under an $SU (1|1)$ rotation, the fields 
$\psi _{{a} } =\textstyle \left(\matrix{\textstyle B_{{a} }\cr F_{{a} } } \right)$
transform as
\begin{eqnarray}\label{psi}
\psi _{{a}}\rarrow e^{A} \psi _{{a}}e^{-A} = \psi_{{a}} +[A,\psi_{{a}}
]+\frac{1}{2}[A, [A, \psi_{{a}} ]]
\end{eqnarray}
where the Grassman coefficients truncate the
expansion at second-order.
Expanding this expression gives
$\psi _{a}\rarrow h \psi _{a}$,
$\psi \dg _{b}\rarrow \psi \dg  _{b}h\dg $
where
\[
h=\left(\matrix{\sqrt{1-\bar \eta \eta  }&-\bar \eta\cr
\eta &\sqrt{1- \eta \bar \eta  } } \right).
\]
is an $SU (1|1)$ matrix, satisfying $h\dg h=1$
The ${X}$-operators (\ref{xdef}) can be written as 
${X}_{{a} {b} }= \psi_{a} \dg \psi_{b}$.
Under the action of the $SU (1|1)$ group, 
$X_{ab}\rarrow \psi\dg  _{a}h\dg h\psi _{b}= X_{ab}  $, explicity
demonstrating the local gauge invariance. 

To guarantee that the Hubbard operator representation is irreducible,
we need to set the values of the linear and quadratic Casimirs of the
$SU (N\vert 1)$ group.
Under the $SU (N|1)$ group, the spinors $B$ and $F$
transform according to $B\rarrow B \tilde{U}$, $F\rarrow F
\tilde{U}$,\cite{note} where $\tilde{U}\equiv U^{st}$ denotes the
supertranspose of the unitary $SU (N\vert 1)$ matrix, $U\ $
\cite{transposeref}.The Hubbard operators $X_{ab}=B\dg _{a}B_{b}+F\dg
_{a}F_{a}$ thus transform according to $X_{ab}\rarrow (\tilde{U}\dg X
\tilde{U})_{ab}$.  Since $U\dg U=1$, it follows that $U^{st} (
U\dg)^{st}=1 $. However, the supertranspose and hermitian conjugate 
do not commute and are related by
$( U\dg)^{st}= g (U^{st})\dg g $, where $g= Diag[1\dots 1,-1]$
is the invariant metric tensor of $SU (N\vert 1)$. Thus 
the $\tilde{U}$ are not unitary, but satisfy
$\tilde{U}g \tilde{U}\dg = g$.  Using the property that
$\hbox{Tr}[AB]=\hbox{Tr}[BgAg]$, it follows that
\begin{equation}\label{casimirs}
C^{( 1)} = \hbox{Tr}[{ X}],\qquad 
C^{( 2)}= \hbox{Tr}[XgX],
\end{equation}
are invariant under the transformation $X\rarrow \tilde{U}\dg X\tilde{U}$. These
are the linear and quadratic Casimirs of the $SU (N|1)$ group. 
Inserting (\ref{therep}) into (\ref{casimirs}), we find that 
$C^{(1)}=Q$,
while the quadratic Casimir is
\begin{equation}\label{quadratic}
C^{( 2)}= X_{\sigma \sigma '}X_{\sigma' \sigma }- X_{\sigma 0}X_{0\sigma }+
X_{0 \sigma '}X_{\sigma '0} - X_{00}^{2}
\end{equation}
where summation over $\sigma, \sigma '\in \{1,N \} $ is implied. When we
expand the Casimir in terms of the canonical creation and annihilation
operators, we find that
\begin{equation}\label{quad2}
C^{(2)}=\hat Q (N-1-{\hat Y}),
\end{equation}
with $Q$ and $Y$ as given in
(\ref{constraint1}) and (\ref{constraint2} ). So by defining ${Y}$ and $Q$, we uniquely set the
representation. 

\begin{figure}
\begin{center}
\includegraphics[scale=0.6]{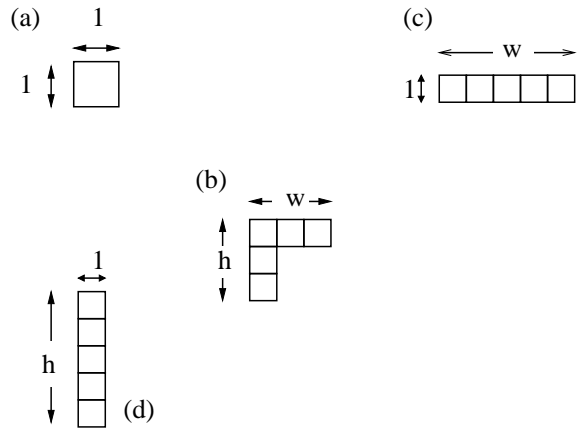}
\vskip .5truein
\caption{\label{two}  (a) Fundamental representation $(Q,Y)= (1,0)$, (b) L-shaped Young tableau 
corresponding to the spin representation generated by  supersymmetric
Hubbard operators.  The asymmetry ${Y}=h-w $ and 
$Q$ is the number of boxes, (c) Young tableau for fully symmetric representation
corresponding to Slave fermion limit (d) Fully antisymmetric, slave boson limit. 
}
\end{center} \end{figure}

Each conserved value of $(Q,Y)$ describes an irreducible
representation of the $SU (N\vert 1)$ group; the fundamental
representation, 
$(Q,Y)=(1,0) $ 
corresponds  to an atomic orbital with no
double occupancy (Fig. \ref{two}(a)).  
More general representations involve spin wavefunctions 
with symmetric and antisymmetric correlations, 
denoted by an ``$L-$ shaped'' Young tableau\cite{tableau} with
$Q$ boxes, where ${Y}=h-w$ is the difference between the height
and width (Fig. \ref{two} (b-d)). 
These representations describe 
the physics of multi-electron atoms
where the spins are Hund's coupled, and in this way
strong Hund's couplings can be incorporated into 
an infinite $U$ Anderson model using the constraints
(\ref{constraint1})
and (\ref{constraint2} ).
As an example, the material $LiV_{2}O_{4}$ develops 
a paramagnetic 
heavy fermion ground-state\cite{liv2O4} in which vanadium ions
form a mixed valence admixture of a 
$d^{1} (S=1/2)$ and a Hund's coupled $d^{2} (S=1) $ state.
Since the electrons in the $d^{{2}}$ configuration are in a symmetric
wavefunction, 
corresponding to a row-tableau, this situation is described 
by Hubbard operators in the representation
$(Q,Y)= (2,-1)$:
\[
{e^{-}+\Yvcentermath1
\overbrace{\yng (1)}^{\hbox{$d^{1}$}}\rightleftharpoons \overbrace
{\yng (2)}^{\hbox{$d^{2}$}} }\quad .
\] 
As a second example,  consider $UPd_{2}Al_{3}$
in which uranium atoms 
fluctuate between an $f^{2}$ and an $f^{{3}}$
configuration. Surprisingly, 
part of the spin
magnetically orders, while the remainder forms a singlet superconductor
with the conduction electrons.\cite{upd2al3}
In this case,  the f-electrons  are spin-orbit coupled, with 
$j=5/2$, forming 
an $SU (N)$
multiplet with $N=2j+1=6$. In practice, crystal field effects
break this large degeneracy, but a toy model
for the physics can be obtained using $SU (N)$ Hubbard operators to
describe the charge fluctuations, subject to the constraint
$(Q,Y)= (2,1)$. This leads to valence
fluctuations involving an $L-$ shaped spin $f^{3}$ spin configuration: 
\[
{e^{-}+\Yvcentermath1
\overbrace{\yng (1,1)}^{\hbox{$f^{2}$}}\rightleftharpoons \overbrace
{\yng (2,1)}^{\hbox{$f^{3}$}} }\quad .
\] 
In this scheme the vertical leg of the representation can form 
a singlet with conduction electrons, leaving a single residual 
spin free to magnetically order.\cite{cpt2}

In many problems we are interested in interacting atoms containing
either one, or zero electrons. Physical states corresponding to this
situation have $Q=1, Y=0$:
\begin{equation}\label{q1y0}
\hat  Q \vert \psi  \ra = \vert \psi \ra, \quad \hat {Y}\vert
\psi \ra =0.
\end{equation}
These conditions do
not force the representation into a slave boson, or
slave fermion representation. Here, it is useful to
note that $\theta $ and $\theta \dg $ behave as lowering, and 
raising operators. In fact, because $\{\theta ,\theta \dg  \}=Q$,
\[
\tau _{+}= \frac{1}{\sqrt{Q}}\theta \dg, \quad \tau _{-}=
\frac{1}{\sqrt{Q}}\theta , \quad 
\tau _{z}= [\tau_{+},\tau_{-}]=\frac{1}{{Q}}[\theta\dg  ,\theta ], 
\]
behave as the raising, lowering 
and z components of a ``superspin'' operator. 
If we take the sum
and difference of the constraints (\ref{constraint1}) and
(\ref{constraint2}), we find that for $Q=1$
\begin{eqnarray}\label{look}
n_{f}+ n_{\phi }&=& \frac{1}{2} (1+\tau _{z}),\cr
n_{b}+ n_{\chi  }&=& \frac{1}{2} (1-\tau _{z}).
\end{eqnarray}
For $\tau _{z}=1$ these equations
revert to the constraints for a slave boson representation, 
when ${\tau_{z}}=-1$, they revert to those of a slave fermion
representation, i.e an ``up'' superspin corresponds to a slave 
boson state, $\frac{1}{2} (1+\tau _{z})\vert \psi \ra = \vert \psi
_{F}\ra$, 
a ``down'' superspin corresponds to a slave-fermion
state $\frac{1}{2} (1-\tau _{z})\vert \psi \ra = \vert \psi
_{B}\ra$. 
In the supersymmetric approach, a partition function of a Hamiltonian $H$, 
involves 
tracing over both slave boson and slave
fermion representations, 
\[
Z=\sum_{\lambda \in F,B}\la \psi _{\lambda }\vert e^{-\beta H}\vert
\psi _{\lambda }\ra.
\]
The trace over both subspaces means that 
the derived path integral has a 
$U (1)\times SU (1\vert 1)$ symmetry and new dynamical degrees
of freedom.
In the slave fermion and slave boson schemes, Fermi liquid and magnetic
phases are manifested as ``Higgs phases'' of the $U (1)$ 
gauge group.  \cite{elitzur}
The enlarged $U (1)\times SU (1\vert 1)$ 
gauge group unifies the slave boson and
slave fermion schemes, but also extends beyond it to furnish a 
potentially wider
class of Higgs phases. 
For instance, 
suppose $H$ is a Hamiltonian, such as the $t-J$ model with both magnetic
and paramagnetic phases, then we expect 
 $\la \tau _{z}\ra =-1 $ in the anti-ferromagnetic (insulating)
ground-state and $\la \tau _{z}\ra =+1$ in the paramagnetic
ground-state, but in addition, there is the
possibility of new saddle-points 
where $\la \tau _{z}\ra $ lies between these two extreme values.

We end with a discussion on  the formulation of the $t-J$
model as a supersymmetric large-$N$ expansion.
To handle anti-ferromagnetic
interactions and electron hopping in a large $N$ expansion, 
we adopt the Read-Sachdev scheme, using Hamiltonians that 
are globally invariant under
the unitary symplectic group
$SP (N)$\cite{readsachdev91}.
This
group is a {\sl subgroup} of $SU (N)$ (defined only for even values of
$N=2n$), so its generators are a subset of the Hubbard operators.
Moreover, 
the groups $SP (2)$ and $SU (2)$ are equivalent. 
In $SP (N)$, 
the spin components are divided
into an equal number of 
``up'' and ``down''  values $\sigma \in (\pm
1, \dots \pm N/2)$; the
unitary matrices of SP(N) 
satisfy 
the condition $U^{T}\underline{\epsilon}U=\underline{\epsilon }$, where 
${\epsilon }_{\sigma \sigma '}= {\rm sgn } (\sigma) \delta _{\sigma
, -\sigma' 
}$. 
The $SP (N)$ $t-J$ model is written\cite{sachdev2000}
\begin{eqnarray}\label{hubbard1}
H&=& -\frac{t}{N} \sum_{(i,j)}[X_{\sigma 0} (i){\rm X}_{0\sigma } (j)+{\rm
H.c}]\cr
&+& \frac{J}{N}\sum_{i,j}\epsilon _{\sigma \sigma
'}\epsilon  _{\eta \eta
'}X_{\sigma \eta '} (i)X_{\sigma '\eta } (j)
-\mu \sum_{j}{\cal  N}_{j}
\end{eqnarray}
where ${\cal  N}_{j}= \sum_{\sigma }X_{\sigma
\sigma } (j)$ is the number of particles. 
In the supersymmetric representation, this model becomes $H+\sum
_{j }K_{j}$
\begin{eqnarray}\label{hubbard2}
H &=& - \frac{t}{N}\sum_{( i,j)}[(f\dg _{i\sigma }\phi _{i}+b\dg _{i\sigma }\chi
_{i})
(\phi \dg _{j}f _{j\sigma }+\chi \dg _{j}b _{j\sigma }) +{\rm H.c. }]\cr
&-&\frac{J}{N} \sum_{( i,j)}
{\rm Tr}[\Lambda\dg _{ij}\Lambda_{ij}]-\mu \sum_{j} {\cal N}_{j},
\end{eqnarray}
where 
$K_{j}=\lambda _{j} ( \hat  Q_{j}-Q_{0}) +\zeta _{j} ( Y_{j}-Y_{0})
$
describes the constraints at site $j$, ${\cal N}_{j}=n_{f} (j)+n_{b}
(j)$ and 
\[
\Lambda_{ij}= {\epsilon }_{\si \si' }\left[\matrix{
f_{i\sigma }f_{j\sigma' }&
f_{i\sigma }b_{j\sigma '}\cr
b_{i\sigma }f_{j\sigma' }&
b_{i\sigma }b_{j\sigma '}
} \right]
\]
describes the singlet valence bonds
between site $i$ and site $j$. 
This Hamiltonian is invariant under the global $SP (N)$ transformation
and the local $U
(1)\times SU (1\vert 1)$ gauge group. The family of models with 
$(Q_{0},Y_{0})= (N/2,0)$, ($N$ even) are of particular interest. 
Two points deserve  special mention:

i) In a path integral treatment, by carrying out a local gauge
transformation $\psi_{j} \rarrow g_{i} (\tau )\psi _{i}$ and
integrating over 
$g_{j}$, one obtains a 
supersymmetric Lagrangian\cite{cpt1}, ${\cal
L}={\cal  L}_{susy}+ H$, where 
\[
{\cal  L}_{susy}=\sum_{j,a}\psi_{ja}\dg \biggl[
\partial_{\tau} - \lambda_{j} + \zeta_{j} \tau_3\biggr]
\psi_{ja} 
- \frac{1}{Q_o} 
\theta_{j}\dg(\partial_{\tau } + 2 \zeta_{j}) \theta_{j} .
\]
This is the starting point for the study of the various
Higgs phases of the model. In each of these phases, 
one of the fermi fields is absorbed into the fluctuations
of the gauge field. 
For instance, in paramagnetic phases the slave boson condenses and
by fixing 
\[
\psi_{j} =g_{j}  \left(\matrix{
b'_{j\sigma_{1}}&\dots  b'_{j\sigma_{N}}
&0\cr 
f'_{j\sigma_{1}}&\dots  f'_{j\sigma_{N}}
& r_{j} } \right),
\]
the slave fermions  $\chi_{j} $ are 
absorbed into the gauge field. Similary, 
the Schwinger boson field $b_{\sigma }$ condenses in an  ordered 
anti-ferromagnetic phase, absorbing a component of the $f_{\sigma }$
fields. More complex Higgs phases, in which fermi fields 
of the bond variables are absorbed into plaquet fermions
also become possible.

ii) The Lagrange multiplier $\zeta _{j}$ which imposes the constraint on 
$Y_{j}$ gives rise to a self-consistently determined spin interaction
 $H_{I}=-\frac{2\zeta_{j}  }{Q}\theta_{j} \dg \theta_{j} $, 
resembling recent approaches to 
the Hubbard model in which spin interactions 
self-consistently renormalize to enforce local constraints\cite{tremblay}.
The Gaussian fluctuations of the $\theta $ fields associated with
this spin interaction 
play a crucial role
in enforcing the constraints between slave boson and slave fermion fields,
and non-trivial results depend on the inclusion 
of these fluctuations in the effective action.

In conclusion, we have presented a 
supersymmetric 
representation of Hubbard operators in which both the operators {\sl
and} the constraints
are invariant under the action of the supergroup  $U
(1)\times SU (1|1)$.
This approach avoids the need to choose between
a fermionic, or bosonic representation for spins.
The underlying $U (1)\times SU (1\vert 1)$ gauge group 
is larger than the
simple product of two $U (1)$ gauge groups. 
Broken symmetry 
saddle points of this enlarged group provide the opportunity to
study the interplay between magnetism and paramagnetism. 

This work was supported  in part by the National Science Foundation
under grants  DMR 9983156 (PC and JH) and PHY 99-07947 (PC and CP) and research funds from the
EPSRC, UK (CP). PC and CP would like to thank the 
Isaac Newton Institute 
and the Institute for Theoretical Physics,
Santa Barbara, where part of this work was carried out.


\begin{references}
\bibitem{note}The transpose matrices must be used, because both $B$ and
$F$ are column supervectors.
\bibitem{elitzur}In these phases, 
Elitzur's theorem assures that 
the local $U (1)$ gauge group is not
actually broken, but the slave fields at each site develop 
power-law correlations in time and may 
in this sense be said to have developed
long-range order in time.
\end{references}
\end{document}